\documentclass{llncs}

\usepackage{amsmath, amssymb, theorem}
\usepackage{stmaryrd}

\newenvironment{ttprog}{\begin{trivlist}\small\tt\item\begin{tabbing}}%
        {\end{tabbing}\end{trivlist}}

\newcommand{\ignore}[1]         {}

\newcommand{\A}                 {\mathcal A}
\newcommand{\B}                 {\mathcal B}
\newcommand{\C}                 {\mathcal C}
\newcommand{\G}                 {\mathcal G}
\newcommand{\V}                 {\mathcal V}
\newcommand{\p}                 {\mathcal P}

\newcommand{\logic}[1]          {\llbracket #1 \rrbracket}
\newcommand{\simparrow}[0]      {\Longleftrightarrow}
\newcommand{\proparrow}[0]      {\Longrightarrow}
\newcommand{\rewritearrow}      {\ \rightarrow\ }
\newcommand{\CHROR}[0]          {CHR$^\vee$}
\newcommand{\where}[0]          {{~{where}~}}
\newcommand{\whereT}[0]         {{where}}
\newcommand{\subterm}[2]        {{{#1}|_{#2}}}
\newcommand{\repl}[3]           {{{#1}[#2]_{#3}}}
\newcommand{\pos}[1]            {{\cal P}os(#1)}

\newcommand{\num}[2]            {#1\##2}

\newcommand{\cc}                {\mathbin{\backslash}}

\newcommand{\lseq}              {\mathit{leq}}

\newcommand{\intg}              {\mathit{int}}
\newcommand{\real}              {\mathit{real}}
\newcommand{\pair}              {\mathit{pair}}

\newcommand{\ltrue}             {\mathsf{true}}
\newcommand{\lfalse}            {\mathsf{false}}

\newcommand{\rewritesto}        {\rightarrowtail}

\begin{document}

\title{ACD Term Rewriting}
\author{Gregory J. Duck \and Peter J. Stuckey \and
        Sebastian Brand}
\institute{
    NICTA Victoria Laboratory\\
    Department of Computer Science \& Software Engineering, \\
    University of Melbourne, Australia
}

\maketitle

\begin{abstract}
In this paper we introduce Associative Commutative Distributive Term
Rewriting (ACDTR), a rewriting language for rewriting logical formulae.
ACDTR extends AC term
rewriting by adding \emph{distribution}
of conjunction over other operators.
Conjunction is vital for expressive term rewriting systems since it allows
us to require that multiple conditions hold
for a term rewriting rule to be
used. ACDTR uses the notion of a ``conjunctive context'',
which is the conjunction of constraints that must
hold in the context of a term, to enable the programmer
to write very expressive and targeted rewriting rules.
ACDTR can be seen as a general logic
programming language that extends Constraint Handling Rules and
AC term rewriting. In this paper we define the semantics of
ACDTR and describe our prototype implementation.
\end{abstract}

\section{Introduction}\label{sec:intro}

Term rewriting is a powerful instrument to
specify computational processes.  It is
the basis of functional languages;
it is used to define the semantics of languages
and it is applied in automated theorem proving,
to name only a few application areas.

One difficulty faced by users of term rewriting systems is that
term rewrite rules are \emph{local}, that is, the
term to be rewritten occurs in a single place.
This means in order to write precise
rewrite rules we need to gather all relevant information in a
single place.

\begin{example}\label{ex:types}
Imagine we wish to ``program'' an overloaded ordering relation
for integers variables, real variables and pair variables.
In order to write this the ``type'' of the variable must be encoded
in the term%
\footnote{Operator precedences used throughout this paper are:
$\land$ binds tighter than $\lor$, and
all other operators, e.g.\ $\lnot$, =, bind tighter than $\land$.}
as in:
\[
\begin{array}{rcl}
int(x) \leq int(y) &\rewritearrow& intleq(int(x),int(y)) \\
real(x) \leq real(y) &\rewritearrow& realleq(real(x),real(y)) \\
pair(x_1,x_2) \leq pair(y_1,y_2) &\rewritearrow& x_1 \leq y_1 \vee x_1 = y_1
\wedge x_2 \leq y_2
\end{array}
\]
In a more standard language, the type information for variables (and other
information) would be kept separate and ``looked up'' when required.
\qed
\end{example}

Term rewriting systems such as constraint handling
rules (CHRs)~\cite{CHR} and
associative commutative (AC) term rewriting~\cite{tr}
allow ``look up'' to be managed straightforwardly
for a single conjunction.

\begin{example}
In AC term rewriting the above example could be expressed as:
\[
\begin{array}{rcl}
\intg(x) \wedge \intg(y) \wedge x \leq y &\rewritearrow& \intg(x) \wedge
\intg(y) \wedge \mathit{intleq}(x,y) \\
\real(x) \wedge \real(y) \wedge x \leq y &\rewritearrow&
\real(x) \wedge \real(y) \wedge \mathit{realleq}(x,y) \\
\pair(x,x_1,x_2) \wedge \pair(y,y_1,y_2) \wedge  x \leq y &\rewritearrow&
\pair(x,x_1,x_2)
\wedge \pair(y,y_1,y_2) \wedge \\
&& (x_1 \leq y_1 \vee x_1 = y_1
\wedge x_2 \leq y_2)
\end{array}
\]
where each rule replaces the $x \leq y$ by an appropriate
specialised version, in the conjunction of constraints.
The associativity and commutativity of $\wedge$ is used to easily collect
the required type information from a conjunction.
\qed
\end{example}

One difficulty remains with both AC term rewriting and CHRs.
The ``look up'' is restricted to be over a single large conjunction.
\begin{example}
Given the term
$\intg(x_1) \wedge \intg(y_1) \wedge
\pair(x,x_1,x_2) \wedge \pair(y,y_1,y_2) \wedge {}$ $x \leq y$.
Then after rewriting $x \leq y$ to $(x_1 \leq y_1 \vee x_1 = y_1
\wedge x_2 \leq y_2)$
we could not rewrite
$x_1 \leq y_1$ since the types for $x_1, y_1$
appear in a different level.

In order to push the type information inside the disjunction we need to
distribute conjunction over disjunction.
\qed
\end{example}
Simply adding distribution rules like
\begin{eqnarray}
A \wedge (B \vee C) &\rewritearrow& A \wedge B \vee A \wedge C \label{rule:dist-expand}\\
A \wedge B \vee A \wedge C &\rewritearrow& A \wedge (B \vee C)
\label{rule:dist-contract}
\end{eqnarray}
does not solve the problem.
Rule~\eqref{rule:dist-expand} creates two copies of term $A$,
which increases the size of the term being rewritten.
Adding Rule~\eqref{rule:dist-contract} to counter this effect
results in a non-terminating rewriting system.

\ignore{
One difficulty faced by users of term rewriting systems
is handling commutativity and associativity (AC).
We quote from Baader and Nipkow~\cite{tr}:
\begin{quote}\small
Probably the most infuriating limitation of the basic term rewriting
framework is the inability to deal gracefully with commutative
operators: commutativity cannot be oriented into a terminating
rewrite rule.
\end{quote}
Term rewriting implementations address this
issue by dealing with AC on the language level
rather than presuming the user to add the actual rewrite rules.
AC function symbols are annotated as such and
term matching is then done under the corresponding AC equality theory.

Distributivity is another property of interest
besides associativity and commutativity.
It causes similar difficulties.
Suppose that the operator $\circ$ is left-distributive
over $\diamond$, that is,
\begin{align}
        \nonumber
        A \circ (B \diamond C) &\approx (A \circ B) \diamond (A \circ C).
\intertext{Its two corresponding rewrite rules are}
        \label{rule:dist-expand}
        A \circ (B \diamond C) &\rewritearrow (A \circ B) \diamond (A \circ C),\\
        \label{rule:dist-contract}
        (A \circ B) \diamond (A \circ C) &\rewritearrow A \circ (B \diamond C).
\end{align}
Rule~\eqref{rule:dist-expand} creates two copies of term $A$,
which increases the size of the term being rewritten.
Adding Rule~\eqref{rule:dist-contract} to counter this effect
results in a non-terminating rewriting system.
}


\subsection{Conjunctive context}

We address the non-termination vs.\ size explosion problem
due to distributivity rewrite rules
in a similar way to how commutativity is dealt with:
by handling distributivity on the language level.
We restrict ourselves to dealing with expanding distributivity
of conjunction $\land$ over any other operator,
and we account for idempotence of conjunction.%
\footnote{
This means that conjunction is distributive over any function $f$ in
presence of a redundant copy of $P$, i.e.
$P \land (P \land f(Q_1, \ldots, Q_n)) \rewritearrow
P \land f(P \land Q_1, \ldots, P \land Q_n)$.
We use idempotence to simplify the RHS and derive
(\ref{rule:conj-dist-expand}).
}
Thus we are concerned with distribution rules of the form
\begin{align}
        \label{rule:conj-dist-expand}
        P \land f(Q_1, \ldots, Q_n) \rewritearrow
        P \land f(P \land Q_1, \ldots, P \land Q_n).
\end{align}

Let us introduce the conjunctive context of a term and its use
in rewrite rules, informally for now.
Consider a term $T$ and the conjunction $\C \land T$
modulo idempotence of $\land$
that would result from exhaustive application of
rule~\eqref{rule:conj-dist-expand} to the superterm of $T$.
By the \emph{conjunctive context} of $T$ we mean the conjunction~$\C$.
\begin{example}\label{ex:linear-small}
The conjunctive context of the boxed occurrence of $x$ in the term
\begin{align*}
        (x = 3) \land ( x^2 > y \ \lor\  (\boxed{x} = 4) \land U \ \lor\  V) \land W,
\end{align*}
is $(x = 3) \land U \land W$.
\qed
\end{example}
We allow a rewrite rule $P \rewritearrow T$ to refer to the
conjunctive context $\C$ of the rule head $P$.
We use the following notation:
\[
        \C \cc P \simparrow T.
\]
This facility provides $\land$-distributivity without the undesirable
effects of rule $\eqref{rule:conj-dist-expand}$ on the term size.
\begin{example}\label{ex:substitution}
We can express that an equality can be used anywhere
``in its scope'' by viewing the equality as a conjunctive context:
\[
        x = a \cc x \simparrow a.
\]
Using this rule on the term of Example~\ref{ex:linear-small} results in
\begin{align*}
        (x = 3) \land ( 3^2 > y \ \lor\ (3 = 4) \land U \ \lor\  V) \land W
\end{align*}
without dissolving the disjunction.
\qed
\end{example}

\subsection{Motivation and Applications}

\subsubsection{Constraint Model Simplification.}

Our concrete motivation behind associative commutative distributive term
rewriting (ACDTR) is \emph{constraint model mapping} as part of
the G12 project~\cite{g12-iclp}. A key aim of G12 is the mapping of
solver independent models to efficient solver dependent models.
\ignore{
The G12 project is made up from three different programming
languages: (1) Zinc a constraint modelling language (2) Cadmium a constraint
model mapping language and (3) Mercury~\cite{mercury} the target CLP
language.
This paper is concerned with Cadmium, which is a language for
mapping solver-independent models to solver-dependent versions.

For example, a solver-independent Zinc model can be mapped to a version
sing a finite domain solver or a different version using a linear solver.
Cadmium is also used for optimising constraint models.

Operationally, Cadmium is closely related to
\emph{Constraint Handling Rules} (CHRs)~\cite{CHR}, which are an
established constraint mapping language.\footnote{
Here \emph{mapping} simply means rewriting constraints into other constraints.
Constraint solving (what CHRs were originally designed for) can be thought of
as a special case of constraint mapping.
}
Cadmium must extend the functionality of CHRs, since CHRs operate on a
\emph{CHR store} (which is essentially a flattened conjunction of constraints),
and constraint models are typically more structured.
}
We see ACDTR as the basis for writing these mappings. Since models
are not flat conjunctions of constraints we need to go beyond AC term
rewriting or  CHRs.

\begin{example}\label{ex:zinc}
Consider the following simple constraint model inspired by the
Social Golfers problem.
For two groups $g_1$ and $g_2$
playing in the same week
there can be no overlap in players: $maxOverlap(g_1,g_2,0)$
The aim is to maximise the
number of times the overlap between two groups is
less than 2; in other words minimise the number of times
two players play together in a group.
\begin{align*}
& \text{constraint}~ \bigwedge_{
        \substack{\forall w \in \mathit{Weeks}\\
                  \forall g_1, g_2 \in \mathit{weeks}[w]\\
                  g_1 < g_2}
    }
    \mathit{maxOverlap}(g_1,g_2,0) \\
& \text{maximise}~~~
\sum_{
    \substack{\forall w_1, w_2 \in \mathit{Weeks}\\
              \forall g_1 \in \mathit{weeks}[w_1]\\
              \forall g_2 \in \mathit{weeks}[w_2]\\
              g_1 < g_2}
    }
    \mathit{holds}(\mathit{maxOverlap}(g_1,g_2,1))
\end{align*}

Consider the following ACDTR program for optimising
this constraint model.
\[
\begin{array}{rcl}
\mathit{maxOverlap}(a,b,c_1) \cc \mathit{maxOverlap}(a,b,c_2) &\simparrow& c_2 \geq c_1 ~|~ \mathit{true} \\
\mathit{holds}(\mathit{true}) &\simparrow& 1 \\
\mathit{holds}(\mathit{false}) &\simparrow& 0 \\
\end{array}
\]
\ignore{
\begin{ttprog}
maxOverlap(A,B,C1) \verb+\+ maxOverlap(A,B,C2) <=> C2 >= C1 | true. \\
holds(true) ~<=> 1. \\
holds(false) <=> 0.
\end{ttprog}
}
The first rule removes redundant $\mathit{maxOverlap}$ constraints.
The next two rules implement partial evaluation of the
$\mathit{holds}$ auxiliary function which coerces a Boolean to an integer.

By representing the constraint model as a giant term, we can
optimise the model by applying the ACDTR program.
For example, consider the trivial case
with one week and two groups $G_1$ and $G_2$.
The model becomes
\[
\mathit{maxOverlap}(G_1,G_2,0) \wedge \mathit{maximise}%
        (\mathit{holds}(\mathit{maxOverlap}(G_1,G_2,1))).
\]
The subterm $\mathit{holds}(\mathit{maxOverlap}(G_1,G_2,1))$ simplifies to
$1$ using the conjunctive context $\mathit{maxOverlap}(G_1,G_2,0)$.
\qed
\end{example}

It is clear that pure CHRs are insufficient for constraint model mapping
for at least two reasons, namely
\begin{itemize}
    \item a constraint model, e.g.\ Example~\ref{ex:zinc}, is typically not
a flattened conjunction;
    \item some rules rewrite functions, e.g. rules (2) and (3) rewriting
    function $\mathit{holds}$, which is outside the scope of CHRs (which
    rewrite constraints only).
\end{itemize}

\subsubsection{Global Definitions.}

As we have seen conjunctive context matching
provides a natural mechanism for
making global information available.
In a constraint model, structured data and constraint definitions
are typically global, i.e.\ on the top level, while
access to the data and the use of a defined constraint is local,
e.g.\ the type information from Example~\ref{ex:types}.
Another example is partial evaluation.
\begin{example}
The solver independent modelling language has support for arrays.
Take a model having an array $a$ of given values.
It could be represented as the top-level term
$\mathit{array}(a, [3, 1, 4, 1, 5, 9, 2, 7])$.
Deeper inside the model, accesses to the array $a$ occur, such as
in the constraint $x > y + \mathit{lookup}(a, 3)$.
The following rules expand such an array lookup:
\begin{align*}
\mathit{array}(A,\mathit{Array}) \cc \mathit{lookup}(A,\mathit{Index}) & \simparrow
    \mathit{list\_element}(\mathit{Array},\mathit{Index}) \\
\mathit{list\_element}([X|\mathit{Xs}],0) & \simparrow X \\
\mathit{list\_element}([X|\mathit{Xs}],N) & \simparrow N > 0 ~|~ \mathit{list\_element}(\mathit{Xs},N-1)
\end{align*}
Referring to the respective array of the lookup expression
via its conjunctive context
allows us to ignore the direct context of the lookup,
i.e. the concrete constraint or expression in which it occurs.
\qed
\end{example}

\ignore{
\begin{example}
The type of a variable is a property that is best stored once,
at or above the highest-level  occurrence of a variable,
rather than with its every occurrence.
The following two rules specialise a constraint
according to the types of the involved variables:
\begin{ttprog}
integer(X) /\verb+\+ integer(Y) \verb+\+ leq(X,Y) <=> fd\_leq(X,Y).\\
real(X)    /\verb+\+ real(Y)    \verb+\+ leq(X,Y) <=> lp\_leq(X,Y).
\end{ttprog}
\qed
\end{example}

\begin{example}
As a final instance of global information that is best accessed
via conjunctive context matching, we mention constraint definitions.
The term
\begin{ttprog}
definition(ordered\_pair((X,Y)) = leq(X, Y)).
\end{ttprog}
in a model defines the constraint \texttt{ordered\_pair(P)}
on a pair \texttt{P} of values.
Depending on the local context, we may sometimes wish to
replace a constraint by its definition:
\begin{ttprog}
definition(ordered\_pair(P) = Def) \verb+\+
\ldots\  ordered\_pair(P) <=> \ldots\ Def.
\end{ttprog}
\qed
\end{example}
}

\subsubsection{Propagation rules.}

When processing a logical formula, it is often useful to be able
to specify that a new formula $Q$ can be derived from an existing
formula $P$ \emph{without consuming} $P$.
In basic term rewriting,
the obvious rule $P \simparrow P \land Q$ causes trivial non-termination.
This issue is recognised in CHRs,
which provide support for inference or \emph{propagation} rules.
We account for this fact and use rules of the form
$P \proparrow Q$ to express such circumstances.
\begin{example}\label{ex:leq}
The following is the classic CHR \texttt{leq} program reimplemented
for ACD term rewriting (we omit the basic rules for logical connectives):
\[
\begin{array}{rcl@{~~~~}l}
        \lseq(X,X) & \simparrow & \mathit{true} & (\mathit{reflexivity}) \\
        \lseq(X,Y) \cc  \lseq(Y,X) & \simparrow & X = Y & (\mathit{antisymmetry}) \\
        \lseq(X,Y) \cc  \lseq(X,Y) & \simparrow & \mathit{true} & (\mathit{idempotence}) \\
        \lseq(X,Y) \wedge \lseq(Y,Z) & \proparrow & \lseq(X,Z) & (\mathit{transitivity})
\end{array}
\]
\ignore{
\begin{ttprog}
        leq(X,X) <=> true. \\
        leq(X,Y) \verb+\+  leq(Y,X) <=> X = Y. \\
        leq(X,Y) \verb+\+  leq(X,Y) <=> true. \\
        leq(X,Y) /\verb+\+ leq(Y,Z) ==> leq(X,Z).
\end{ttprog}
}
These rules are almost the same as the CHR version,
with the exception of the second and third rule
(\emph{antisymmetry} and \emph{idempotence}) which generalise
its original by using conjunctive context matching.
\qed
\end{example}
Propagation rules are also used for adding redundant information during
model mapping.

The rest of the paper is organised as follows.
Section~\ref{sec:prelim} covers the standard syntax and notation of
term rewriting.
Section~\ref{sec:semantics} defines the declarative and operational semantics
of ACDTR.
Section~\ref{sec:implementation} describes a prototype implementation of
ACDTR as part of the G12 project.
Section~\ref{sec:related} compares ACDTR with related languages.
Finally, in Section~\ref{sec:summary} we conclude.

\section{Preliminaries} \label{sec:prelim}

In this section we briefly introduce the notation and terminology used in
this paper.
Much of this is borrowed from term rewriting~\cite{tr}.

We use ${\cal T}(\Sigma,X)$ to represent the set of all terms
constructed from a set of function symbols $\Sigma$ and set of variables
$X$ (assumed to be countably infinite).
We use $\Sigma^{(n)} \subseteq \Sigma$ to represent the set of function symbols
of arity $n$.

A \emph{position} is a string (sequence) of integers that uniquely determines
a subterm of a term $T$, where $\epsilon$ represents the empty string.
We define function $\subterm{T}{p}$, which returns the subterm of $T$
at position $p$ as
\[
\begin{array}{rl}
    \subterm{T}{\epsilon} & = T \\
    \subterm{f(T_1,\ldots,T_i,\ldots,T_n)}{ip} & = \subterm{T_i}{p}
\end{array}
\]
We similarly define a function $\repl{T}{S}{p}$ which replaces the subterm
of $T$ at position $p$ with term $S$.
We define the set $\pos{T}$ to represent the set of all \emph{positions}
of subterms in $T$.

An \emph{identity} is a pair
$(s,t) \in {\cal T}(\Sigma,X) \times {\cal T}(\Sigma,X)$,
which is usually written as $s \approx t$.
Given a set of identities $E$, we define $\approx_E$ to be the set of
identities closed under the axioms of \emph{equational logic}~\cite{tr},
i.e. symmetry, transitivity, etc.

We define the congruence class
$[T]_{\approx_E} = \{S \in {\cal T}(\Sigma,X) | S \approx_E T\}$
as the set of terms equal to $T$ with respect to $E$.

Finally, we define function $vars(T)$ to return the set of variables in
$T$.

\section{Syntax and Semantics} \label{sec:semantics}

The syntax of ACDTR closely resembles that of CHRs.
There are three types of rules of the following form:
\[
\begin{array}{ll}
{\mbox {\it (simplification)~~~~}} &
r ~@~ H ~\simparrow~  g ~|~ B \\
{\mbox {\it (propagation)}} &
r ~@~ H ~\proparrow~  g ~|~ B \\
{\mbox {\it (simpagation)}} &
r ~@~ C ~\backslash~ H
        ~\simparrow~  g ~|~ B
\end{array}
\]
where $r$ is a \emph{rule identifier}, and \emph{head} $H$,
\emph{conjunctive context} $C$, \emph{guard} $g$
and \emph{body} $B$ are arbitrary terms.
The rule identifier is assumed to uniquely determine the rule.
A program $P$ is a set of rules.

We assume that $vars(g) \subseteq vars(H)$ or
$vars(g) \subseteq vars(H) \cup vars(C)$ (for simpagation rules).
The rule identifier can be omitted.
If $g = true$ then the guard can be omitted.

We present the declarative semantics of ACDTR based
on equational logic.

First we define the set of operators that ACDTR treats specially.
\begin{definition}[Operators]
We define the set of \emph{associate commutative} operators as $AC$.
The set $AC$ must satisfy
    $AC \subseteq \Sigma^{(2)}$ and
    $(\wedge) \in AC$.
\end{definition}
For our examples we assume that $AC = \{\wedge,\vee,+,\times\}$.
We also treat the operator $\wedge$ as \emph{distributive} as explained
below.

ACDTR supports a simple form of guards.
\begin{definition}[Guards]
A \emph{guard} is a term.
We denote the set of all ``true'' guards as $\G$, i.e.
a guard $g$ is said to \emph{hold} iff $g \in \G$.
We assume that $\mathit{true} \in {\cal G}$ and $\mathit{false} \not\in \G$.
\end{definition}

We can now define the declarative semantics for ACDTR.
In order to do so we employ a special binary operator $\whereT$ to
explicitly attach a conjunctive context to a term.
Intuitively, the meaning of $T \where C$ is equivalent to that of
$T$ provided $C$ is $true$,
otherwise the meaning of $T \where C$ is unconstrained.
For Boolean expressions, it is useful to interpret $\whereT$ as conjunction
$\wedge$, therefore $\whereT$-distribution, i.e. identity~(\ref{eq:dist}) below,
becomes equivalent to $\wedge$-distribution \eqref{rule:conj-dist-expand}.
The advantage of distinguishing $\whereT$ and $\wedge$ is that we
are not forced to extend the definition of $\wedge$ to arbitrary
(non-Boolean) functions.

We denote  by $\B$ the following set of \emph{built-in} identities:
{\setcounter{equation}{0}
\begin{align}
A \circ B & \approx B \circ A \label{eq:comm} \\
(A \circ B) \circ C & \approx A \circ (B \circ C) \label{eq:assoc}\\
T & \approx (T \where true) \label{eq:make_true} \\
A \wedge B & \approx (A \where B) \wedge B \label{eq:make_cc} \\
T \where (W_1 \wedge W_2) & \approx (T \where W_1) \where W_2
    \label{eq:comb_cc} \\
f(A_1,...,A_i,...,A_n) \where W & \approx
    f(A_1,...,A_i \where W,...,A_n) \where W \label{eq:dist}
\end{align}
}
for all $\circ \in AC$, functions $f \in \Sigma^{(n)}$, and
$i \in \{1, \ldots, n\}$.
\begin{definition}[Declarative Semantics for ACDTR]\label{def:ds_cad}
The \emph{declarative semantics} for an ACDTR program $P$ (represented as a
multiset of rules) is given by the
function $\logic{}$ defined as follows:
\[
\begin{array}{ll}
\logic{P}                         & =
    \{\logic{\theta(R)} ~|~ \forall R, \theta ~.~
        R \in P \wedge \theta({\sf guard}(R)) \in \G\} \cup \B \\
\logic{H \simparrow g~|~B}            & = \exists_{vars(B) - vars(H)}
        (H \approx B) \\
\logic{C \cc H \simparrow g~|~B} & = \exists_{vars(B) - vars(C,H)}
        (H {~where~} C \approx B {~where~} C) \\
\logic{H \proparrow g~|~B}            & = \exists_{vars(B) - vars(H)}
        (H \approx H \wedge B) \\
\end{array}
\]
where function ${\sf guard}(R)$ returns the guard of a rule.
\end{definition}
The function $\logic{}$ maps ACDTR rules to identities between the head and the
body terms,
where body-only variables are existentially quantified.%
\footnote{
All other variables are implicitly universally quantified, where
the universal quantifiers appear outside the existential ones.
}
Note that there is a new identity for each possible binding of
${\sf guard}(R)$ that holds in $\G$.
A propagation rule is equivalent to a simplification rule that (re)introduces
the head $H$ (in conjunction with the body $B$) in the RHS.
This is analogous to propagation rules under CHRs.

A simpagation rule is equivalent to a simplification rule provided the
conjunctive context is satisfied.

The built-in rules $\B$ from Definition~\ref{def:ds_cad} contain identities for
creating/destroying (\ref{eq:make_true}) and (\ref{eq:make_cc}),
combining/splitting (\ref{eq:comb_cc}), and distributing downwards/upwards
(\ref{eq:dist}) a conjunctive context in terms of the $\where$ operator.

The set $\B$ also contains
identities (\ref{eq:comm}) and (\ref{eq:assoc}) for the
associative/commutative properties of the $AC$ operators.

\begin{example}
Consider the following ACDTR rule and the corresponding identity.
\begin{align} \label{eq:rule}
\logic{X = Y \cc X \simparrow Y} ~~~=~~~
    (Y \where X = Y) \approx (X \where X = Y)
\end{align}
Under this identity and using the rules in $\B$, we can show that
$f(A) \wedge (A = B) \approx f(B) \wedge (A = B)$, as follows.
\[
\begin{array}[b]{lrll}
f(A) \wedge (A = B)                                 &
    \approx_{(\ref{eq:make_cc})} \\
(f(A) \where (A = B)) \wedge (A = B)                &
    \approx_{(\ref{eq:dist})} \\
(f(A \where (A = B)) \where (A = B)) \wedge (A = B) &
    \approx_{(\ref{eq:rule})} \\
(f(B \where (A = B)) \where (A = B)) \wedge (A = B) &
    \approx_{(\ref{eq:dist})} \\
(f(B) \where (A = B)) \wedge (A = B)                &
    \approx_{(\ref{eq:make_cc})} \\
f(B) \wedge (A = B)
\end{array}
\]
\qed
\end{example}

\subsection{Operational Semantics}

In this section we describe the operational semantics of ACDTR.
It is based on the theoretical operational semantics of
CHRs~\cite{chropsem,refined_semantics}.
This includes support for identifiers and propagation histories,
and conjunctive context matching for simpagation rules.

\subsubsection{Propagation history.}

The CHR concept of a \emph{propagation history}, which prevents trivial
non-termination of propagation rules, needs to be generalised over arbitrary
terms for ACDTR.
A propagation history is essentially a record of all propagation rule
applications, which is checked to ensure a propagation rule is not applied
twice to the same (sub)term.

In CHRs, each constraint is associated with a unique \emph{identifier}.
If multiple copies of the same constraint appear in the CHR store, then
each copy is assigned a different identifier.
We extend the notion of identifiers to arbitrary terms.
\begin{definition}[Identifiers]\label{def:identifiers}
An \emph{identifier} is an integer associated with each (sub)term.
We use the notation $\num{T}{i}$ to indicate that term $T$ has been
associated with identifier $i$.
A term $T$ is \emph{annotated} if $T$ and all subterms of $T$ are associated
with an identifier.
We also define function ${\sf ids}(T)$ to return the set of identifiers in
$T$, and ${\sf term}(T)$ to return the non-annotated version of $T$.
\end{definition}
For example, $T = \num{f(\num{a}{1},\num{b}{2})}{3}$ is an annotated term,
where ${\sf ids}(T) = \{1,2,3\}$ and ${\sf term}(T) = f(a,b)$.

Identifiers are considered separate from the
term.
We could be more precise by separating the two, i.e. explicitly maintain
a map between $\pos{T}$ and the identifiers for $T$.
We do not use this approach for space reasons.
We extend and overload all of the standard operations over terms
(e.g. from Section~\ref{sec:prelim}) to annotated terms in the obvious
manner.
For example, the subterm relation $\subterm{T}{p}$ over annotated terms
returns the annotated term at position $p$.
The exception are elements of the congruence class $[T]_{\approx_{AC}}$,
formed by the $AC$ relation $\approx_{AC}$,
which we assume satisfies the following constraints.
\[
\begin{array}{rl}
\num{A}{i} \circ \num{B}{j} & \approx_{AC}
        \num{B}{j} \circ \num{A}{i} \\
\num{A}{i} \circ (\num{B}{j} \circ \num{C}{k}) & \approx_{AC}
        (\num{A}{i} \circ \num{B}{j}) \circ \num{C}{k} \\
\end{array}
\]
We have neglected to mention the identifiers over $AC$ operators.
These identifiers will be ignored later, so we leave them unconstrained.

A propagation history is a set of entries defined as follows.
\begin{definition}[Entries]
A \emph{propagation history entry} is of the form $(r~@~E)$, where
$r$ is a propagation rule identifier, and $E$ is a string of identifiers.
We define function ${\sf entry}(r,T)$ to return the propagation history entry
of rule $r$ for annotated term $T$ as follows.
\[
\begin{array}{lll}
{\sf entry}(r,T)           & = (r~@~{\sf entry}(T)) \\
{\sf entry}(T_1 \circ T_2) & = {\sf entry}(T_1)~{\sf entry}(T_2)
        ~~~~~~~~~ & \circ \in AC \\
{\sf entry}(\num{f(T_1,...,T_n)}{i}) & =
            i~{\sf entry}(T_1)~...~{\sf entry}(T_n) &
            \text{otherwise}
\end{array}
\]
\end{definition}
This definition means that propagation history entries are unaffected by
associativity, but are effected by commutativity.
\begin{example}
Consider the annotated term
$T = \num{f(\num{(\num{a}{1} \wedge \num{b}{2})}{3})}{4}$.
We have that $T \in [T]_{\approx_{AC}}$ and
$T' = \num{f(\num{(\num{b}{2} \wedge \num{a}{1})}{3})}{4} \in
    [T]_{\approx_{AC}}$.
Although $T$ and $T'$ belong to $[T]_{\approx_{AC}}$ they have different
propagation history entries, e.g.
\ignore{
\[
\begin{array}{ll}
{\sf entry}(r,T)  & = (r~@~(4~1~2)) \\
{\sf entry}(r,T') & = (r~@~(4~2~1)) \\
\end{array}
\]}
${\sf entry}(r,T)   = (r~@~(4~1~2))$ while ${\sf entry}(r,T')  = (r~@~(4~2~1))$.
\qed
\end{example}

When a (sub)term is rewritten into another, the new term is assigned
a set of new unique identifiers.
We define the auxiliary function ${\sf annotate}(\p,T) = T_a$ to map a
set of identifiers $\p$ and un-annotated term $T$ to an annotated term
$T_a$ such that ${\sf ids}(T_a) \cap \p = \emptyset$ and
$|{\sf ids}(T_a)| = |\pos{T}|$.
These conditions ensure that all identifiers are new and unique.

When a rule is applied the propagation history must be updated
accordingly to reflect which terms are copied from the matching.
For example, the rule
$f(X) \simparrow g(X,X)$
essentially clones the term matching $X$.
The identifiers, however, are not cloned.
If a term is cloned, we expect that both copies will inherit the
propagation history of the original.
Likewise, terms can be merged, e.g.
$g(X,X) \simparrow f(X)$
merges two instances of the term matching $X$.
In this case, the propagation histories of the copies are also merged.

To achieve this we duplicate entries in the propagation history for each
occurrence of a variable in the body that also appeared in the head.
\begin{definition}[Updating History]\label{def:update}
Define function \[{\sf update}(H,H_a,B,B_a,T_0) = T_1\]
where $H$ and $B$ are un-annotated terms, $H_a$ and $B_a$ are
annotated terms, and $T_0$ and $T_1$ are propagation histories.
$T_1$ is a minimal propagation history satisfying the following conditions:
\begin{itemize}
    \item $T_0 \subseteq T_1$;
    \item $\forall p \in \pos{H}$ such that
    $\subterm{H}{p} = V \in X$ (where $X$ is the set of variables), and
    $\exists q \in \pos{B}$ such that
    $\subterm{B}{q} = V$, then define identifier renaming $\rho$ such that
    $\rho(\subterm{H_a}{p})$ and $\subterm{B_a}{q}$ are identical
    annotated terms.
    Then if $E \in T_0$ we have that $\rho(E) \in T_1$.
\end{itemize}
\end{definition}

\begin{example}
Consider rewriting the term
$H_a = \num{f(\num{(\num{a}{1} \wedge \num{b}{2})}{3})}{4}$
with a propagation history of
$T_0 = \{(r ~@~ (1~ 2))\}$ using the rule $f(X) \simparrow g(X,X)$.
The resulting term is
$B_a = \num{g(\num{(\num{a}{5} \wedge \num{b}{6})}{7}),
              \num{(\num{a}{8} \wedge \num{b}{9})}{10}}{11}
$
and the new propagation history is
$T_1 = \{(r ~@~ (1~ 2)), (r ~@~ (5~ 6)), (r ~@~ (8~ 9))\}$.
\qed
\end{example}

\subsubsection{Conjunctive context.}

According to the declarative semantics, a term $T$ with conjunctive
context $C$ is represented as $(T \where C)$.
Operationally, we will never explicitly build a term containing a
$\whereT$ clause.
Instead we use the following function to compute the conjunctive context
of a subterm on demand.
\begin{definition}[Conjunctive Context]
Given an (annotated) term $T$ and a position $p \in \pos{T}$, we define
function ${\sf cc}(T,p)$ to return the conjunctive context
at position $p$ as follows.
\[
\begin{array}{lll}
    {\sf cc}(T,\epsilon)    & = true \\
    {\sf cc}(A \wedge B,1p) & = B \wedge {\sf cc}(A,p) \\
    {\sf cc}(A \wedge B,2p) & = A \wedge {\sf cc}(B,p) \\
    {\sf cc}(f(T_1,\ldots,T_i,\ldots,T_n),ip)
        & =
        {\sf cc}(T_i,p) ~~~~ ~~~~ ~~~~ ~~~~ ~~~~ ~~~~ & (f \neq \wedge)
\end{array}
\]
\end{definition}

\subsubsection{States and transitions.}

The operational semantics are defined as a set of transitions on
execution states.
\begin{definition}[Execution States]
An \emph{execution state} is a tuple of the form
$\langle G, T, \V, \p \rangle$,
where $G$ is a term (the \emph{goal}), $T$ is the \emph{propagation
history}, $\V$ is the set of variables appearing in the initial goal and
$\p$ is a set of identifiers.
\end{definition}
We also define initial and final states as follows.
\begin{definition}[Initial and Final States]
Given an initial goal $G$ for program $P$, the \emph{initial state} of $G$ is
\[
    \langle G_a, \emptyset, vars(G), {\sf ids}(G_a) \rangle
\]
where $G_a = {\sf annotate}(\emptyset,G)$.
A \emph{final state} is a state where no more rules are applicable to the
goal $G$.
\end{definition}

We can now define the operational semantics of ACDTR as follows.
\begin{definition}[Operational Semantics]
\[
\langle G_0, T_0, \V, \p_0 \rangle \rightarrowtail
    \langle G_1, T_1, \V, \p_1 \rangle
\]
\noindent\textbf{1. Simplify:}
There exists a (renamed) rule from $P$
\[
    H \simparrow g ~|~ B
\]
such that there exists a matching substitution $\theta$ and a term
$G'_0$ such that
\begin{itemize}
    \item $G_0 \approx_{AC} G'_0$
    \item $\exists p \in \pos{G'_0} ~.~ \subterm{G'_0}{p} = \theta(H)$
    \item $\theta(g) \in \G$
    \item $B_a = {\sf annotate}(\p_0,\theta(B))$
\end{itemize}
Then $G_1 = \repl{G'_0}{B_a}{p}$,
$\p_1 = \p_0 \cup {\sf ids}(G_1)$ and
$T_1 = {\sf update}(H,\subterm{G'_0}{p},B,B_a,T_0)$.
\\[1ex]
\noindent\textbf{2. Propagate:}
There exists a (renamed) rule from $P$
\[
    r ~@~ H \proparrow g ~|~ B
\]
such that there exists a matching substitution $\theta$ and a term
$G'_0$ such that
\begin{itemize}
    \item $G_0 \approx_{AC} G'_0$
    \item $\exists p \in \pos{G'_0} ~.~ \subterm{G'_0}{p} = \theta(H)$
    \item $\theta(g) \in \G$
    \item ${\sf entry}(r,\subterm{G'_0}{p}) \not\in T_0$
    \item $B_a = {\sf annotate}(\p_0,\theta(B))$
\end{itemize}
Then $G_1 = \repl{G'_0}{\subterm{G'_0}{p} \wedge B_a}{p}$,
$T_1 = {\sf update}(H,\subterm{G'_0}{p},B,B_a,T_0)
\cup \{{\sf entry}(r,\subterm{G'_0}{p})\}$ and
$\p_1 = \p_0 \cup {\sf ids}(G_1)$.
\\[1ex]
\noindent\textbf{3. Simpagate:}
There exists a (renamed) rule from $P$
\[
    C ~\backslash~ H \simparrow g ~|~ B
\]
such that there exists a matching substitution $\theta$ and a term
$G'_0$ such that
\begin{itemize}
    \item $G_0 \approx_{AC} G'_0$
    \item $\exists p \in \pos{G'_0} ~.~ \subterm{G'_0}{p} =
        \theta(H)$
    \item $\exists D. \theta(C) \wedge D  \approx_{AC} {\sf cc}(G'_0,p)$
    \item $\theta(g) \in \G$
    \item $B_a = {\sf annotate}(\p_0,\theta(B))$
\end{itemize}
Then $G_1 = \repl{G'_0}{B_a}{p}$,
$T_1 = {\sf update}(H,\subterm{G'_0}{p},B,B_a,T_0)$ and
$\p_1 = \p_0 \cup {\sf ids}(G_1)$.
\end{definition}

\subsubsection{Example.}

Consider the $leq$ program from Example~\ref{ex:leq} with the goal
\[
    leq(X,Y) \wedge leq(Y,Z) \wedge \neg leq(X,Z)
\]
Figure~\ref{fig:leq_deriv} shows one possible derivation of this goal
to the final state representing $false$.
For brevity, we omit the $\V$ and $\p$ fields, and represent
identifiers as subscripts, i.e. $\num{T}{i} = T_i$.
Also we substitute $T = \{{\tt transitivity}~@~(3~2~1~7~5~6)\}$.
\begin{figure}[t]
\[
\begin{array}{l}
\langle (leq(X_1,Y_2)_3 \wedge_4 leq(Y_5,Z_6)_7 \wedge_8 \neg_9
leq(X_{10},Z_{11})_{12}),\emptyset \rangle
\rightarrowtail_{trans} \\
\langle (leq(X_1,Y_2)_3 \wedge_4 leq(Y_5,Z_6)_7 \wedge_{13}
    leq(X_{15},Z_{16})_{14} \wedge_8 \neg_9 leq(X_{10},Z_{11})_{12}),
    T \rangle
\rightarrowtail_{idemp} \\
\langle (leq(X_1,Y_2)_3 \wedge_4 leq(Y_5,Z_6)_7 \wedge_{13}
    leq(X_{15},Z_{16})_{14} \wedge_8 \neg_9 true_{17}),
    T \rangle
\rightarrowtail_{simplify} \\
\langle (leq(X_1,Y_2)_3 \wedge_4 leq(Y_5,Z_6)_7 \wedge_{13}
    leq(X_{15},Z_{16})_{14} \wedge_8 false_{18}),
    T \rangle
\rightarrowtail_{simplify} \\
\langle (leq(X_1,Y_2)_3 \wedge_4 leq(Y_5,Z_6)_7 \wedge_{13} false_{19}),
    T \rangle
\rightarrowtail_{simplify} \\
\langle (leq(X_1,Y_2)_3 \wedge_4 false_{20}),
    T \rangle
\rightarrowtail_{simplify} \\
\langle (false_{21}), T \rangle \\
\end{array}
\]
\vspace*{-8mm}
\caption{Example derivation for the $leq$
  program.\label{fig:leq_deriv}}
\vspace*{3mm}
\end{figure}


We can state a soundness result for ACDTR.
\begin{theorem}[Soundness]
If
$\langle G_0, T_0, \V, \p \rangle \rightarrowtail^* \langle G', T', \V, \p' \rangle$
with respect to a program $P$, then
$\logic{P} \models \exists_{vars(G') - \V} ~ G_0 \approx G'$
\end{theorem}
This means that for all algebras $\A$ that satisfy $\logic{P}$,
$G_0$ and $G'$ are equivalent for some assignment of the fresh variables in
$G'$.

\section{Implementation} \label{sec:implementation}

We have implemented a prototype version of ACDTR as part of the
mapping language of the G12 project, called Cadmium.
In this section we give an overview of the implementation details.
In particular, we will focus on the implementation of conjunctive context
matching, which is the main contribution of this paper.

Cadmium constructs \emph{normalised} terms from the bottom up.
Here, a \emph{normalised} term is one that cannot be reduced further
by an application of a rule.
Given a goal $f(t_1,...,t_n)$, we first must
recursively normalise all of $t_1,...,t_n$ (to say $s_1,...,s_n$), and then
attempt to find a rule that can be applied to the top-level of
$f(s_1,...,s_n)$.
This is the standard execution algorithm used by many TRSs
implementations.

This approach of normalising terms bottom up is complicated
by the consideration of conjunctive context matching.
This is because the conjunctive context of the current term
appears ``higher up'' in the overall goal term.
Thus conjunctive context must be passed top down, yet we are normalising
bottom up.
This means there is no guarantee that the conjunctive context is
normalised.
\begin{example}\label{ex:cc_change}
Consider the following ACDTR program that uses conjunctive context
matching.
\[
\begin{array}{rcl}
X = V \cc X &\simparrow& var(X) \wedge nonvar(V) ~|~ V. \\
one(X) &\simparrow& X = 1. \\
not\_one(1) &\simparrow& false.
\end{array}
\]
Consider the goal $not\_one(A) \wedge one(A)$, which we
expect should be normalised to $false$.
Assume that the sub-term $not\_one(A)$
is selected for normalisation
first. The conjunctive context for $not\_one(A)$
(and its subterm $A$) is $one(A)$.
No rule is applicable, so $not\_one(A)$ is not reduced.

Next the subterm $one(A)$ is reduced.
The second rule will fire resulting in the new term $A = 1$.
Now the conjunctive context for the first term $not\_one(A)$ has changed to
$A = 1$, so we expect that $A$ should be rewritten to
the number $1$.
However $not\_one(A)$ has already being considered for normalisation.
\qed
\end{example}

The current Cadmium prototype solves this problem by re-normalising terms
when and if the conjunctive context ``changes''.
For example, when the conjunctive context $one(A)$ changes to
$A = 1$, the term $not\_one(X)$ will be renormalised
to $not\_one(1)$ by the first rule.

The general execution algorithm for Cadmium is shown in
Figure~\ref{fig:cad_exec}.
Function \textsf{normalise} takes a term $T$, a substitution $\theta$,
a conjunctive context $CC$ and a Boolean value $Ch$ which keeps track of
when the conjunctive context of the current subterm has changed.
If $Ch = true$, then we can assume the substitution $\theta$ maps variables to
normalised terms.
For the initial goal, we assume $\theta$ is empty,
otherwise if we are executing a body of a rule, then $\theta$ is the
matching substitution.

Operationally,
\textsf{normalise} splits into three cases depending on what
$T$ is.
If $T$ is a variable, and the conjunctive context has changed
(i.e. $Ch = true$), then $\theta(T)$ is no longer guaranteed to be
normalised.
In this case we return the result of renormalising $\theta(T)$ with respect to
$CC$.
Otherwise if $Ch = false$, we simply return $\theta(T)$ which must be already
normalised.
If $T$ is a conjunction $T_1 \wedge T_2$, we repeatedly call \textsf{normalise}
on each conjunct with the other added to the conjunctive context.
This is repeated until a fixed point (i.e. further normalisation
does not result in either conjunct changing) is reached,
and then return the result of
\textsf{apply\_rule} on the which we will discuss below.
This fixed point calculation accounts for the case where the conjunctive
context of a term changes, as shown in Example~\ref{ex:cc_change}.
Otherwise, if $T$ is any other term of the form $f(T_1,...,T_n)$,
construct the new term $T'$ by normalising each argument.
Finally we return the result of \textsf{apply\_rule} applied to $T'$.

The function call \textsf{apply\_rule}($T'$,$CC$) will attempt to apply a rule
to normalised term $T'$ with respect to conjunctive context $CC$.
If a matching rule is found, then the result of
\textsf{normalise}($B$,$\theta$,$CC$,$false$) is returned, where $B$ is the
(renamed) rule body and $\theta$ is the matching substitution.
Otherwise, $T'$ is simply returned.

\begin{figure}[t]
\begin{tabbing}
xx \= xx \= xx \= xx \= xx \= xx \= xx \= xx \=\kill
\textsf{normalise}($T$,$\theta$,$CC$,$Ch$) \\
\> \textbf{if} $is\_var(T)$ \\
\> \> \textbf{if} $Ch$ \\
\> \> \> \textbf{return} \textsf{normalise}($\theta(T)$,$\theta$,$CC$,$false$)
    \\
\> \> \textbf{else} \\
\> \> \> \textbf{return} $\theta(T)$ \\
\> \textbf{else} \textbf{if} $T = T_1 \wedge T_2$ \\
\> \> \textbf{do} \\
\>\>\> $T_1'$ := $T_1$ \\
\>\>\> $T_2'$ := $T_2$ \\
\>\>\> $T_1$ :=
    \textsf{normalise}($T_1'$,$\theta$,$T_2' \wedge CC$,$true$) \\
\>\>\> $T_2$ :=
    \textsf{normalise}($T_2'$,$\theta$,$T_1' \wedge CC$,$true$) \\

\>\> \textbf{while} $T_1 \neq T_1' \wedge T_2 \neq T_2'$ \\
\> \> \textbf{return} \textsf{apply\_rule}($T_1' \wedge T_2'$,$CC$) \\
\> \textbf{else} \\
\> \> $T$ = $f(T_1,...,T_n)$ \\
\> \> $T'$ := $f($\textsf{normalise}($T_1$,$\theta$,$CC$,$Ch$)$,...,$
                \textsf{normalise}($T_n$,$\theta$,$CC$,$Ch$)$)$ \\
\> \textbf{return} \textsf{apply\_rule}($T'$,$CC$)
\end{tabbing}
\vspace*{-5mm}
\caption{Pseudo code of the Cadmium execution
  algorithm.\label{fig:cad_exec}}
\vspace*{3mm}
\end{figure}

\section{Related Work} \label{sec:related}

ACDTR is closely related to both TRS and CHRs, and in this section
we compare the three languages.

\subsection{AC Term Rewriting Systems}

The problem of dealing with associative commutative operators in TRS
is well studied.
A popular solution is to perform the rewriting modulo some permutation of the
AC operators.
Although this complicates the matching algorithm, the problem of
trivial non-termination (e.g. by continually rewriting with respect to
commutativity) is solved.

ACDTR subsumes ACTRS (Associative Commutative TRS) in that we have
introduced distributivity (via simpagation rules), and added some
``CHR-style'' concepts such as identifiers and propagation rules.

Given an ACTRS program, we can map it to an equivalent ACDTR program
by interpreting each ACTRS rule $H \rightarrow B$ as the ACDTR rule $H \simparrow B$.
We can now state the theorem relating ACTRS and ACDTR.
\begin{theorem}
Let $P$ be an ACTRS program and $T$ a ground term, then
\mbox{$T \rightarrow^* S$} under $P$ iff
$\langle T_a, \emptyset, \emptyset, {\sf ids}(T_a) \rangle \rightarrowtail^*
 \langle S_a, \emptyset, \emptyset, \p \rangle$ under $\alpha(P)$
(where $T_a = {\sf annotate}(\emptyset,T)$) for some $\p$
and ${\sf term}(S_a) = S$.
\end{theorem}

\subsection{CHRs and \CHROR}

ACDTR has been deliberately designed to be an extension of CHRs.
Several CHR concepts, e.g. propagation rules, etc., have been adapted.

There are differences between CHRs and ACDTR.
The main difference is that ACDTR does not have
a ``built-in'' or ``underlying'' solver, i.e. ACDTR is not a constraint
programming language.
However it is possible to encode solvers directly as rules, e.g.
the simple $leq$ solver from Example~\ref{ex:leq}.
Another important difference is that CHRs is based on predicate logic,
where there exists a distinction between predicate symbols
(i.e. the names of the constraints) and functions
(used to construct terms).
ACDTR is based on equational logic between terms, hence there is no
distinction between predicates and functions
(a predicate is just a Boolean function).
To overcome this, we assume the existence of a set ${\cal P}red$,
which contains the set of function symbols that are Boolean functions.
We assume that $AC \cap {\cal P}red = \{\wedge^{(2)}\}$.

The mapping between a CHR program and an ACDTR program is
simply $\alpha(P) = P \cup \{X \wedge true \simparrow X\}$.\footnote{
There is one slight difference in syntax: CHRs use `\texttt{,}' to
represent conjunction, whereas ACDTR uses `$\wedge$'.
}
However, we assume program $P$ is restricted as follows:
\begin{itemize}
    \item rules have no guards apart from implicit equality guards; and
    \item the only built-in constraint is $true$
\end{itemize}
and the initial goal $G$ is also restricted:
\begin{itemize}
    \item $G$ must be of the form
    $G_0 \wedge ... \wedge G_n$ for $n > 0$;
    \item Each $G_i$ is of the form $f_i(A_0,...,A_m)$ for $m \geq 0$ and
    $f_i \in {\cal P}red$;
    \item For all $p \in \pos{A_j}, 0 \leq j \leq m$ we have that
    if $\subterm{A_j}{p} = g(B_0,...,B_q)$ then $g^{(q)} \not\in AC$
    and $g^{(q)} \not\in {\cal P}red$.
\end{itemize}
These conditions disallow predicate symbols from appearing as arguments
in CHR constraints.

\begin{theorem}\label{th:chrs}
Let $P$ be a CHR program, and $G$ an initial goal both satisfying the
above conditions, then
$\langle G, \emptyset, true, \emptyset \rangle^\V_1 \rightarrowtail
    \langle \emptyset, S, true, T \rangle^\V_i$
(for some $T$, $i$ and $\V = vars(G)$) under the theoretical operational
semantics~\cite{refined_semantics} for CHRs
iff
$\langle G_a, \emptyset, \V, {\sf ids}(G_a) \rangle
    \rightarrowtail
 \langle S_a, T', \V, \p \rangle$ (for some $T'$, $\p$) under ACDTR,
 where ${\sf term}(S_a) = S_1 \wedge ... \wedge S_n$ and
 $S = \{\num{S_1}{i_1},...,\num{S_n}{i_n}\}$
 for some identifiers $i_1,...,i_n$.
\end{theorem}

We believe that Theorem~\ref{th:chrs} could be extended to include
CHR programs that extend an underlying solver, provided the rules for
handling tell constraints are added to the ACDTR program.
For example, we can combine rules for rational tree unification
with the $\lseq$ program from
Example~\ref{ex:leq} to get a program equivalent to the traditional
$\lseq$ program under CHRs.

ACDTR generalises CHRs by allowing other operators besides conjunction
inside the head or body of rules.
One such extension of CHRs has been studied before, namely
\CHROR~\cite{chror} which allows disjunction in the body.
Unlike ACDTR, which manipulates disjunction syntactically, \CHROR
typically finds solutions using backtracking search.

One notable implementation of \CHROR is~\cite{java_chror}, which has
an operational semantics described as an and/or ($\wedge$/$\vee$)
tree rewriting system.
A limited form of conjunctive context matching is used, similar to that
used by ACDTR, based on the knowledge
that conjunction $\wedge$ distributes over disjunction $\vee$.
ACDTR generalises this by distributing over all functions.

\section{Future Work and Conclusions} \label{sec:summary}

We have presented a powerful new rule-based programming language, ACDTR,
that naturally extends both AC term rewriting
and CHRs.
The main contribution is the ability to match a rule against the
conjunctive context of a (sub)term, taking advantage of the
distributive property of conjunction over all possible functions.
We have shown this is a natural way of expressing some problems,
and by building the distributive property into the matching algorithm,
we avoid non-termination issues that arise from naively implementing
distribution (e.g. as rewrite rules).

We intend that ACDTR will become the theoretical basis for the
Cadmium constraint mapping language as part of the G12
project~\cite{g12-iclp}.
Work on ACDTR and Cadmium is ongoing, and there is a wide scope for future
work, such as confluence, termination and implementation/optimisation
issues.


\bibliographystyle{plain}



\newpage
\appendix
\section{Examples}

\subsection{Further Motivating Examples}\label{sec:examples}

\begin{example}[Conjunctive Normal Form]\label{ex:cnf}
One of the roles of mapping models is to convert a model
written in an expressive language into a restricted language
which is easy to solve.
Many standard approaches to solving propositional formulae
require that the formulae are in conjunctive normal form (CNF).
Disjunction $\lor$ is distributive over $\land$,
which can be used to establish CNF in a direct way,
using the oriented rule
\[
        P \lor Q \land R \rewritearrow (P \lor Q) \land (P \lor R).
\]
CNF conversion based on this rule can
exponentially increase the size of the formula.
This undesirable circumstance means that in practice
CNF conversions are preferred that replace
subformulae by new propositional atoms,
which increases the formula size at most linearly.

Let us formulate this approach in rewrite rules.
To keep this example simple, we assume that the non-CNF subformula
$P \lor Q \land R$ occurs in a positive context (for example by a
preprocessing into negation normal form).
We replace $Q \land R$ by a new atom $s$
defined by the logical implication $s \Rightarrow (Q \land R)$.
In rewrite rule form, we have
\begin{align}
        P \lor Q \land R & \rewritearrow (P \lor s) \land (\lnot s \lor Q) \land (\lnot s \lor R).
\end{align}
Unit resolution and unit subsumption can be formalised in rewrite rules.
Here are two versions, one using conjunctive context and
a regular one:
\[\begin{array}{@{}ll}
\begin{split}
\text{with conj.\ context:}                     \\[1ex]
P \cc P & \simparrow \ltrue                     \\\\
P \cc \lnot P & \simparrow \lfalse              \\\\
\end{split}
&\hspace{2em}
\begin{split}
\text{regular:}                                 \\[1ex]
        P \land P & \rewritearrow P             \\
        P \land (P \lor Q) & \rewritearrow P    \\
        P \land \lnot P & \rewritearrow \lfalse   \\
        P \land (\lnot P \lor Q) & \rewritearrow P \land Q
\end{split}
\end{array}\]
We furthermore assume rules eliminating
the logical constants $\ltrue$ and $\lfalse$ from conjunctions and disjunctions
in the obvious way.
Let us contrast the two rule sets for the formula
$(a \lor b \land (c \lor d)) \land d$.  The following is a
terminating rewrite history:
\[\begin{array}{@{}ll}
\begin{split}
&\text{with conj.\ context:}\\[1ex]
&
(a \lor b \land (c \lor d)) \land d             \\\quad\rewritesto\quad&
(a \lor b \land (c \lor \ltrue)) \land d        \\\quad\rewritesto\quad&
(a \lor b \land \ltrue) \land d                 \\\quad\rewritesto\quad&
(a \lor b) \land d
\end{split}
&\hspace{3em}
\begin{split}
&\text{regular:}\\[1ex]
&
(a \lor b \land (c \lor d)) \land d                                     \\\quad\rewritesto\quad&
(a \lor s) \land (\lnot s \lor b) \land (\lnot s \lor c \lor d) \land d \\\quad\rewritesto\quad&
(a \lor s) \land (\lnot s \lor b) \land \ltrue \land d                  \\\quad\rewritesto\quad&
(a \lor s) \land (\lnot s \lor b) \land d
\end{split}
\end{array}\]
To obtain the simple conjunct $(a \lor b)$
using the regular rule format,
a rule expressing binary resolution, i.e.\
from $(P \lor S) \land (\lnot S \lor Q)$ follows $(P \lor Q)$,
would be required.  However, such a rule is undesirable
as it would create arbitrary binary resolvents,
increasing formula size.
Moreover, the superfluous atom $s$ remains in the formula.
\qed
\end{example}

\begin{example}[Type remapping]
One of the main model mappings we are interested in expressing is
where the type of a variable is changed from a high level type
easy for modelling to a low level type easy to solve.
A prime example of this is mapping a set variable $x$ ranging
over finite subsets of some fixed set $s$ to
an array $x'$ of 0/1 variables indexed by $s$.
So for variable $x$ we have $e \in x \Leftrightarrow x'[e] = 1$.
For this example we use the more concrete modelling syntax:
$t:x$ indicates variable $x$ has type $t$, the types we are interested are
$l..u$ an integers in the range $l$ to $u$,
$set~of~S$ a set ranging over elements in $S$, and $array[I]~of~E$ an
array indexed by set $I$ of elements of type $E$.
We use $forall$ and $sum$ looping constructs
which iterate over sets.
This is expressed in ACDTR as follows.
\[
\begin{array}{rcll}
set~of~s:x &\simparrow& array[s] ~of~ 0..1:x' \wedge map(x,x') & (typec) \\
map(x,x') \cc x &\simparrow& x' & (vsubs) \\
array[s]~ of~ 0..1:x \cc card(x) &\simparrow& sum(e ~in~ s)~ x[e] & (card) \\
\begin{array}{r}
array[s]~ of~ 0..1:x ~\wedge \\
array[s]~ of~ 0..1:y \cc x \cap y
\end{array}
&\simparrow &
\begin{array}{l}
z :: (array[s]~ of~ 0..1:z ~\wedge \\
forall(e ~in~ s)~ z[e] = x[e] ~\&\&~ y[e])
\end{array}
& (cap) \\
\begin{array}{r}
array[s]~ of~ 0..1:x ~\wedge \\
array[s]~ of~ 0..1:y \cc x \cup y
\end{array}
&\simparrow &
\begin{array}{l}
z :: (array[s]~ of~ 0..1:z ~\wedge \\
forall(e ~in~ s)~ z[e] = x[e] ~||~ y[e])
\end{array}
& (cup) \\
array[s]~ of~ 0..1:x \cc
 x = \emptyset  & \simparrow &
forall(e ~in~ s)~ x[e] = 0 & (emptyset)\\
card(t :: c) &\simparrow & card(t) :: c &(\uparrow{}card) \\
(t_1 :: c) \cup t_2 & \simparrow & t_1 \cup t_2 :: c&(\uparrow{}cupl) \\
t_1 \cup (t_2 :: c) & \simparrow & t_1 \cup t_2 :: c&(\uparrow{}cupr) \\
(t_1 :: c) \cap t_2 & \simparrow & t_1 \cap t_2 :: c&(\uparrow{}capl) \\
t_1 \cap (t_2 :: c) & \simparrow & t_1 \cap t_2 :: c&(\uparrow{}capr) \\
(t_1 :: c) = t_2 & \simparrow & t_1 = t_2 \wedge c&(\uparrow{}eql) \\
t_1 = (t_2 :: c) & \simparrow & t_1 = t_2 \wedge c&(\uparrow{}eqr) \\
(t_1 :: c) \leq t_2 & \simparrow & t_1 \leq t_2 \wedge c&(\uparrow{}leql) \\
t_1 \leq (t_2 :: c) & \simparrow & t_1 \leq t_2 \wedge c&(\uparrow{}leqr) \\
(t :: c_1) :: c_2 & \simparrow & t :: (c_1 \wedge c_2)&(\uparrow{}cc) \\
maxOverlap(x,y,c) & \simparrow & card(x \cap y) \leq c & (maxO) \\
\end{array}
\]
The $::$ constructor adds some local conjunctive context to an
arbitrary term (like $where$) and the last 11 rules bar 1 move this
context outwards to the nearest predicate scope. The last rule
defines the maxOverlap predicate.
They are used to introduce new variables $z$ and their type
and the constraints upon then.
As an example, consider the following derivation:
 \[
\begin{array}{r@{~~~~}l}
& \underline{set~ of~ 1..n:x} \wedge set~ of~ 1..n:y \wedge maxOverlap(x,y,1)\\
\rightarrowtail_{maxO} & \underline{set~ of~ 1..n:x} \wedge set~ of~ 1..n:y \wedge card(x \cap y) \leq 1  \\
\rightarrowtail_{typec} & array[1..n]~of~0..1:x' \wedge \underline{map(x,x')}
\wedge set~ of~ 1..n:y \wedge card(\underline{x} \cap y) \leq 1  \\
\rightarrowtail_{vsubs} & array[1..n]~of~0..1:x' \wedge map(x,x')
\wedge \underline{set~ of~ 1..n:y} \wedge card(x' \cap y) \leq 1  \\
\rightarrowtail_{typec} & array[1..n]~of~0..1:x' \wedge map(x,x')
\wedge array[1..n]~of~0..1:y' \wedge \underline{map(y,y')} ~\wedge \\
& card(x' \cap \underline{y}) \leq 1  \\
\rightarrowtail_{vsubs} & array[1..n]~of~0..1:x' \wedge map(x,x')
\wedge array[1..n]~of~0..1:y' \wedge map(y,y') ~\wedge \\
& card(\underline{x' \cap y'}) \leq 1  \\
\rightarrowtail_{cap} & array[1..n]~of~0..1:x' \wedge map(x,x')
\wedge array[1..n]~of~0..1:y' \wedge map(y,y') ~\wedge \\
& \underline{card(z :: (array[1..n]~of~0..1:z \wedge forall(e ~in~
1..n)~ z[e] = x'[e] ~\&\&~ y'[e])} \leq 1 \\
\rightarrowtail_{\uparrow{}card} & array[1..n]~of~0..1:x' \wedge map(x,x')
\wedge array[1..n]~of~0..1:y' \wedge map(y,y') ~\wedge \\
& \underline{card(z) :: (array[1..n]~of~0..1:z \wedge forall(e ~in~
1..n)~ z[e] = x'[e] ~\&\&~ y'[e]) \leq 1} \\
\rightarrowtail_{\uparrow{}leql} & array[1..n]~of~0..1:x' \wedge map(x,x')
\wedge array[1..n]~of~0..1:y' \wedge map(y,y') ~\wedge \\
& card(z) \leq 1 \wedge array[1..n]~of~0..1:z \wedge forall(e ~in~
1..n)~ z[e] = x'[e] ~\&\&~ y'[e] \\
\end{array}
\]
The final goal is a flat conjunction of constraints and types.
It can be similarly translated into a
conjunction of pseudo-Boolean constraints that
can be sent to a finite domain solver,
by unrolling $forall$ and replacing
the arrays by sequences of $n$ variables.
\qed
\end{example}

\begin{example}[Rational Tree Unification]\label{ex:unification}
We can directly express the rational tree unification
algorithm
of Colmerauer\footnote{A. Colmerauer. Prolog and Infinite Trees.
\emph{Logic Programming, APIC Studies in Data Processing (16)}. Academic
Press. 1992} as
an ACD term rewriting system.
\[
\begin{array}{rcl@{~~~~}l}
f(s_1, \ldots s_n) = f(t_1, \ldots, t_n) &\simparrow& s_1 = t_1 \wedge
\cdots s_n = t_n & (split) \\
f(s_1, \ldots s_n) = g(t_1, \ldots, t_m) &\simparrow& false & (fail)
\end{array}
\]
The (split) rule must be defined for each constructor $f/n$
and the (fail)
rule for each pair of different constructors $f/n$ and $g/m$.\
The remaining rules are:
\[
\begin{array}{rcl@{~~~~}l}
x = x &\simparrow& var(x) ~|~ true  & (id) \\
t = x &\simparrow& var(x) \wedge nonvar(t) ~|~ x = t &(flip) \\
x = s \cc x = t &\simparrow& var(x) \wedge nonvar(s) \wedge size(s)
\leq size(t) ~|~ s = t & (tsubs) \\
x = y \cc x &\simparrow& var(x) \wedge var(y) \wedge x \not\equiv y ~|~ y & (vsubs)
\end{array}
\]
\ignore{
\begin{ttprog}
split @ f(S1, ..., Sn) = f(T1, ..., Tn) <=> S1 = T1 /\verb+\+ ... /\verb+\+ Sn = Tn \\
fail  @ f(S1, ..., Sn) = g(T1, ..., Tm) <=> false \\
id    @ X = X <=> var(X) | true \\
flip  @ T = X <=> var(X) /\verb+\+  nonvar(T) | X = T \\
tsubs @ X = S \verb+\+ X = T <=> X != S /\verb+\+ size(T) >= size(S) | S = T \\
vsubs @ X = Y \verb+\+ X <=> var(X) /\verb+\+ var(Y) | Y
\end{ttprog}
}
where  $size(t)$ is the size of the term $t$ in terms of number of symbols,
and $\equiv$ is syntactic identity.
Even though the goals are a single conjunction of constraints,
ACD is used for succinctly expressing
the (vsubs) rule which replaces one variable
by another in any other position.

The following derivation illustrates the unification process in action.
The underlined part show the matching elements
\[
\begin{array}{r@{~~~~}l}
& x = y \wedge \underline{f(f(x)) = x} \wedge y = f(f(f(y))) \\
\rightarrowtail_{flip} & \underline{x = y} \wedge \underline{x} = f(f(x)) \wedge y = f(f(f(y))) \\
\rightarrowtail_{vsubs} & \underline{x = y} \wedge y = f(f(\underline{x}))
\wedge y = f(f(f(y))) \\
\rightarrowtail_{vsubs} & x = y \wedge \underline{y = f(f(y)) \wedge y = f(f(f(y)))} \\
\rightarrowtail_{tsubs} & x = y \wedge y = f(f(y)) \wedge \underline{f(f(y))
  = f(f(f(y)))} \\\rightarrowtail_{split} & x = y \wedge y = f(f(y)) \wedge
\underline{f(y) = f(f(y))} \\
\rightarrowtail_{split} & x = y \wedge \underline{y = f(f(y)) \wedge y = f(y)} \\
\rightarrowtail_{tsubs} & x = y \wedge \underline{f(y) = f(f(y))} \wedge y = f(y) \\
\rightarrowtail_{split} & x = y \wedge \underline{y = f(y) \wedge y = f(y)} \\
\rightarrowtail_{tsubs} & x = y \wedge y = f(y) \wedge \underline{f(y) = f(y)} \\
\rightarrowtail_{split} & x = y \wedge y = f(y) \wedge \underline{f(y) = f(y)} \\
\rightarrowtail_{id} & x = y \wedge y = f(y) \wedge true \\
\end{array}
\]
\qed
\end{example}

\subsection{Expanded Examples}\label{sec:expanded}

The purpose of this section is to show some example derivations under the
operational semantics of ACDTR, rather than high-level descriptions.
We allow for some shorthand, namely $\num{T}{i} = T_i$.

\subsubsection{Identifiers and conjunctive context.}

In this section we explain parts of the derivation from
Example~\ref{ex:unification} in more detail.
The initial goal is
\[
x = y \wedge f(f(x)) = x \wedge y = f(f(f(y)))
\]
which corresponds to the initial state:
\[
\begin{array}{c}
\langle (((x_1 = y_2)_3 \wedge (f(f(x_4)_5)_6 = x_7)_8)_9
    \wedge (y_{10} = f(f(f(y_{11})_{12})_{13})_{14})_{15})_{16},
    \emptyset, \\
~~~~~~~~~~~~~ \{x,y\},\{1,2,3,4,5,6,7,8,9,10,11,12,13,14,15,16\}\rangle
\end{array}
\]
The initial state is a quadruple contained an annotated version of the
goal, an empty propagation history, the set of variables in the goal
and a set of ``used'' identifiers.

The first derivation step is a \textbf{Simplify} transition with
the $flip$ rule:
\[
\begin{array}{c}
\langle (((x_1 = y_2)_3 \wedge \underline{(f(f(x_4)_5)_6 = x_7)_8})_9
    \wedge (y_{10} = f(f(f(y_{11})_{12})_{13})_{14})_{15})_{16},
    \emptyset, \\
~~~~~~~~~~~~~ \{x,y\},\{1,2,3,4,5,6,7,8,9,10,11,12,13,14,15,16\}\rangle \\
    \rightarrowtail \\
\langle (((x_1 = y_2)_3 \wedge
    (x_{17} = f(f(x_{18})_{19})_{20})_{21})_9
    \wedge (y_{10} = f(f(f(y_{11})_{12})_{13})_{14})_{15})_{16},
    \emptyset, \\
~~~~~~~~~~~~~ \{x,y\},\{1,2,3,4,5,6,7,8,9,10,11,12,13,14,15,16,17,18,19,20,21\}
    \rangle \\
\end{array}
\]
We have replaced the annotated subterm $(f(f(x_4)_5)_6 = x_7)_8$
with $x_{17} = f(f(x_{18})_{19})_{20})_{21}$ (i.e. flipped the operands
to the equality) and reannotated the new term with fresh identifiers.
These were also added to the set of used identifiers.
Since the propagation history is empty, it remains unchanged.

The next derivation step is a \emph{Simpagate} transition with the
$vsubs$ rule.
\[
\begin{array}{c}
\langle ((\underline{(x_1 = y_2)_3} \wedge
    (\underline{x_{17}} = f(f(x_{18})_{19})_{20})_{21})_9
    \wedge (y_{10} = f(f(f(y_{11})_{12})_{13})_{14})_{15})_{16},
    \emptyset, \\
~~~~~~~~~~~~~ \{x,y\},\{1,2,3,4,5,6,7,8,9,10,11,12,13,14,15,16,17,18,19,20,21\}
    \rangle \\
    \rightarrowtail \\
\langle (((x_1 = y_2)_3 \wedge
    (y_{21} = f(f(x_{18})_{19})_{20})_{21})_9
    \wedge (y_{10} = f(f(f(y_{11})_{12})_{13})_{14})_{15})_{16},
    \emptyset, \\
~~~~~~~~~~~~~ \{x,y\},
        \{1,2,3,4,5,6,7,8,9,10,11,12,13,14,15,16,17,18,19,20,21,22\}
    \rangle \\
\end{array}
\]
The conjunctive context for subterm $x_{17}$ is
\[
    {\sf cc}(G_a,p) = (x_1 = y_2)_3 \wedge
    (y_{10} = f(f(f(y_{11})_{12})_{13})_{14})_{15} \wedge
    true
\]
where $G_a$ is the current goal and $p$ is the position of $x_{17}$.
The first conjunct matches the conjunctive context of the $vsubs$ rule,
thus subterm $x_{17}$ is replaced with $y_{21}$.
Identifier $21$ is added to the list of used identifiers.

Execution proceeds until the final state
\[
    \langle  (x = y \wedge y = f(y)) \wedge true, \emptyset, \{x,y\},
        \p \rangle
\]
is reached, for some annotation of the goal and some set of identifiers $\p$.
This is a final state because no more rules are applicable to it.

\subsubsection{AC matching and propagation histories.}

Consider the propagation rule from the $\lseq$ program:
\[
trans~@~\lseq(X,Y) \wedge \lseq(Y,Z) \proparrow X \not\equiv Y
    \wedge Y \not\equiv Z ~|~ \lseq(X,Z)
\]
and the initial state
\[
\langle \lseq(A_1,B_2)_3 \wedge_4 \lseq(B_5,A_6)_7, \emptyset, \{A,B\},
    \{1,2,3,4,5,6,7\} \rangle.
\]

We can apply \textbf{Propagate} directly (i.e. without permuting the
conjunction) to arrive at the state:
\[
\begin{array}{c}
\langle (\lseq(A_1,B_2)_3 \wedge_4 \lseq(B_5,A_6)_7)
    \wedge_8 \lseq(A_9,A_{10})_{11}, \\
        \{trans~@~(3~1~2~7~6~5)\},
        \{1,2,3,4,5,6,7,8,9,10,11\} \rangle.
\end{array}
\]

The propagation history prevents the rule from firing on the same
terms again, however we can permute the terms to find a new matching.
Namely, we can permute the annotated goal (which we call $G_a$)
\[
(\lseq(A_1,B_2)_3 \wedge_4 \lseq(B_5,A_6)_7) \wedge_8 \lseq(A_9,A_{10})_{11}
\]
to
\[
(\lseq(B_5,A_6)_7 \wedge_4 \lseq(A_1,B_2)_3) \wedge_8 \lseq(A_9,A_{10})_{11}.
\]
The latter is an element of $[G_a]_{AC}$, and the identifiers have been preserved
in the correct way.
The entry $trans~@~(7~6~5~3~1~2)$ is not in the propagation history, so we can
apply \textbf{Propagate} again to arrive at:
\[
\begin{array}{c}
\langle ((\lseq(B_5,A_6)_7 \wedge_4 \lseq(A_1,B_2)_3) \wedge_{12}
    \lseq(B_{13},B_{14})_{15}) \wedge_8 \lseq(A_9,A_{10})_{11}, \\
        \{trans~@~(3~1~2~7~6~5), trans~@~(7~6~5~3~1~2)\},
        \{1...15\} \rangle.
\end{array}
\]

Now the propagation history prevents the rule $trans$ being applied to
the first two $\lseq$ constraints.
The guard also prevents the $trans$ rule firing on either of the two new
constraints,\footnote{
Without the guard both ACDTR and CHRs are not guaranteed to terminate.
}
thus we have reached a final state.

\subsubsection{Updating propagation histories.}

Consider a modified version of the previous example, now with two rules:
\begin{align*}
    X \wedge X & \simparrow X \\
    trans ~@~ \lseq(X,Y) \wedge \lseq(Y,Z) & \proparrow \lseq(X,Z)
\end{align*}
The first rule enforces \emph{idempotence} of conjunction.

Consider the initial state:
\[
\begin{array}{c}
\langle \lseq(A_1,A_2)_3 \wedge_4 \lseq(A_5,A_6)_7 \wedge_8
    \lseq(A_9,A_{10})_{11}, \emptyset,
    \{A\}, \{1...11\} \rangle
\end{array}
\]
We apply the $trans$ rule to the first two copies of the $\lseq$ constraint
(with identifiers 3 and 7).
\[
\begin{array}{c}
\langle \lseq(A_1,A_2)_3 \wedge_4 \lseq(A_5,A_6)_7 \wedge_8
    \lseq(A_9,A_{10})_{11} \wedge_{12} \lseq(A_{13},A_{14})_{15}, \\
    \{trans~@~(3~1~2~7~5~6)\}, \{A\}, \{1...15\} \rangle
\end{array}
\]
Next we apply idempotence to $\lseq$ constraints with identifiers
7 and 11.
\[
\begin{array}{c}
\langle \lseq(A_1,A_2)_3 \wedge_4 \lseq(A_{16},A_{17})_{18}
    \wedge_{12} \lseq(A_{13},A_{14})_{15}, \\
    \{trans~@~(3~1~2~7~5~6),trans~@~(3~1~2~18~16~17)\}, \{A\}, \{1...18\}
\rangle
\end{array}
\]
An extra entry $(trans~@~(3~1~2~18~16~17))$ is added to the propagation history
in order to satisfy the requirements of Definition~\ref{def:update}.
This is because we have replaced the annotated constraint
$\lseq(A_5,A_6)_7$ with the newly annotated term $\lseq(A_{16},A_{17})_{18}$,
which defines an identifier renaming
\[
\rho = \{5 \mapsto 16, 6 \mapsto 17, 7 \mapsto 18\}.
\]
Since $E = (trans~@~(3~1~2~7~5~6))$ is an element of the propagation history,
we have that $\rho(E) = (trans~@~(3~1~2~18~16~17))$ must also be an element,
and hence the history is expanded.

\end{document}